\documentclass[12pt]{amsart}
\usepackage{amsfonts,amssymb,amsmath,amscd}
\usepackage{epsf}

\newcommand{\bj}{{\bar{j}}}
\newcommand{\bi}{{\bar{i}}}

\renewcommand{\d}{{\rm d}}

\newcommand{\om}{\omega}

\def\part{\partial}

\newcommand{\ii}{{\sqrt{-1}}}

\newcommand{\CC}{{\mathbb C}}
\newcommand{\PP}{{\mathbb P}}
\newcommand{\RR}{{\mathbb R}}
\newcommand{\ZZ}{{\mathbb Z}}

\newcommand{\QQ}{{\mathbb Q}}

\newcommand{\hT}{{\hat T}}
\newcommand{\hG}{{\hat G}}
\newcommand{\hB}{{\hat B}}
\newcommand{\hP}{{\hat P}}
\newcommand{\bG}{{\bar G}}
\newcommand{\Thol}{{T_{\rm hol}}}
\newcommand{\I}{{\mathfrak I}}

\newcommand{\tI}{{\tilde I}}
\newcommand{\tP}{{\tilde P}}
\newcommand{\tom}{{\tilde\omega}}

\newcommand{\End}{{\rm End}}

\newcommand{\cA}{{\mathcal A}}

\newcommand{\cG}{{\mathcal G}}
\newcommand{\cI}{{\mathcal I}}
\newcommand{\cJ}{{\mathcal J}}

\renewcommand{\O}{{\mathcal O}}

\newcommand{\ra}{\rightarrow}
\newcommand{\bpartial}{{\bar\partial}}
\newcommand{\no}{\nonumber}
\newcommand{\op}{\oplus}
\newcommand{\ot}{\otimes}

\newcommand{\hp}{\hbar}
\newcommand{\ch}{{\rm ch}}

\newcommand{\Lie}{{\mathcal L}}
\newcommand{\hx}{{\hat x}}
\newcommand{\hI}{{\hat I}}
\newcommand{\htheta}{{\hat\theta}}
\newcommand{\hA}{{\hat A}}
\newcommand{\hF}{{\hat F}}

\newcommand{\bE}{{\bar E}}
\newcommand{\cE}{{\mathcal E}}
\newcommand{\Ker}{{\rm Ker}}
\newcommand{\cP}{{\mathcal P}}
\renewcommand{\i}{\iota}
\newcommand{\V}{{\mathcal V}}
\newcommand{\bV}{{\bar{\mathcal V}}}

\title[Topological strings on noncommutative manifolds]{Topological Strings on Noncommutative Manifolds}
\author[A. Kapustin]{Anton Kapustin \hspace{3mm} }
\address{California Institute of Technology\\
Department of Physics\\
Pasadena, CA 91125}
\email{kapustin@theory.caltech.edu}

\date{August 2003}

\begin{document}

\begin{abstract}
We identify a deformation of the $N=2$ supersymmetric sigma model on a Calabi-Yau manifold $X$
which has the same effect on B-branes as a noncommutative deformation of $X$. We show that for
hyperk\"ahler $X$ such deformations allow one to interpolate continuously between the A-model and the B-model.
For generic values of the noncommutativity and the B-field, properties of the topologically twisted
sigma-models can be described in terms of generalized complex structures introduced by N. Hitchin.
For example, we show that the path integral for the deformed sigma-model is localized on generalized
holomorphic maps, whereas for the A-model and the B-model it is localized on holomorphic and constant maps,
respectively. The geometry of topological D-branes is also best described using generalized complex
structures. We also derive a constraint on the Chern character of topological D-branes, which includes
A-branes and B-branes as special cases.

\end{abstract}

\maketitle

\vspace{-5in}

\parbox{\linewidth}
{\small\hfill \shortstack{CALT-68-2457}} 

\vspace{5in}

\section{Introduction}

One interesting recent development in string theory is a realization of the role of noncommutative
geometry. For example, it has been shown that a certain limit of string theory in flat space-time is described by field
theory on a noncommutative affine space. This limit is a version of the ``zero-slope'' limit $\alpha'\ra 0$, with the B-field
held constant~\cite{SW}. 
To see the emergence of noncommutativity, it is necessary to study open strings (i.e. strings ending on
D-branes), since the zero-slope limit of the closed string CFT always gives a commutative algebra. 

It is not
well understood how to extend these considerations to more general manifolds. In this paper we address this issue
in the context of topological string theory. Topological string amplitudes have simpler dependence on $\alpha'$, 
so one might hope to see the emergence of noncommutative geometry without taking an elaborate limit.

Topological strings on a Calabi-Yau manifold $X$ come in two flavors~\cite{Witten}.
The A-model and the corresponding D-branes (A-branes)
depend only on the symplectic structure and the B-field on $X$, while the B-model and B-branes depend only on the complex
structure and the $(0,2)$ part of the B-field. The category of B-branes on $X$
is believed to be equivalent to the bounded derived category of coherent sheaves on $X$, denoted $D^b(X)$~\cite{Douglas}
(see Ref.~\cite{Sharpe} for a review).
Objects of this category can be thought of as complexes of holomorphic vector bundles on $X$. In turn, a holomorphic vector 
bundle can be thought of as a locally free sheaf of modules over the sheaf $\O_X$ of holomorphic functions on $X$. 
Deforming the sheaf $\O_X$ into a sheaf of noncommutative algebras, one gets the corresponding deformation
of the category of B-branes. As far as we know, A-branes can not be thought in terms of 
modules over a commutative algebra, even locally, so it does not make much sense to talk about noncommutative deformations of 
A-branes.

Our starting point is the mathematical classification of infinitesimal deformations of the category of B-branes on $X$,
regarded as an $A_\infty$ category. According to Ref.~\cite{Konts1}, the space of infinitesimal deformations
is the Hochschild cohomology of $D^b(X)$, which is isomorphic to
$$
\op_{p,q} H^p(\Lambda^q \Thol X).
$$
We will be interested in deformations which are marginal in the sense of world-sheet CFT. This means that we will look at the
component of this vector space with $p+q=2$. 

The piece $H^1(\Thol X)$ classifies infinitesimal deformations of the complex structure on $X$.
These are ``commutative'' deformations of the category of B-branes. It is well-known that for Calabi-Yau manifolds
the deformation problem is unobstructed, in the sense that any infinitesimal deformation can be ``exponentiated''
to an actual deformation.

The other two pieces are more interesting.\footnote{They are present only if $h^{0,2}(X)\neq 0$, in which case
the holonomy of $X$ is strictly smaller than $SU(n)$, i.e. it admits more than one covariantly constant spinor.}
Consider first the component with $p=2,q=0$, i.e. $H^2(\O)=H^{0,2}(X).$ Deformations of this kind correspond
to replacing coherent sheaves with twisted coherent sheaves. Let us recall what this means (see Refs.~\cite{KB,CKS} and
references therein for
more details). Let $\lambda\in H^2(\O)$ be the deformation of interest, and let us choose an open cover $U_i,$ $i\in \I,$
such that $\lambda$ can be represented by a Cech cocycle $\lambda_{ijk}$. We will denote by $U_{i_1\ldots i_k}$
the open set $U_{i_1}\cap\ldots\cap U_{i_k}$. A $\lambda$-twisted coherent sheaf on $X$ is
a collection of coherent sheaves $S_i$ on $U_i$ and sheaf isomorphisms
$$
\phi_{ji}:S_i\vert_{U_{ij}}\ra S_j\vert_{U_{ij}}
$$
such that 
\begin{itemize}
\item[(i)] $\phi_{ii}=id$ for all $i\in\I$,
\item[(ii)] $\phi_{ji}=\phi_{ij}^{-1}$ for all $i,j\in\I$,
\item[(iii)] $\phi_{ij}\phi_{jk}\phi_{ki}=\exp(\ii\lambda_{ikj})$ for all $i,j,k$.
\end{itemize}
Twisted coherent sheaves form an abelian category, and one can define its derived category in the usual manner.
The bounded derived category of $\lambda$-twisted coherent sheaved is the deformation of $D^b(X)$ in the
direction $\lambda$.

From the physical viewpoint, every deformation of the category of B-branes should correspond to some deformation of
the topologically twisted sigma-model on a world-sheet with boundaries. In the case of $H^2(\O)$ there is
an obvious candidate: the $(0,2)$ part of the B-field. It is not hard to see that for $B^{0,2}\neq 0$ to any B-brane
one can associate a twisted coherent sheaf. Indeed, suppose we are dealing with a D-brane which is a vector bundle $E$ on $X$;
all other D-branes can be obtained by considering complexes (i.e. bound states) of such branes. 
For $B=0$ the condition of BRST-invariance reads $F^{0,2}=0$, where $F$ is the curvature 2-form of the connection $\nabla$
on $E$. Recall that the effect of a flat B-field on D-branes is to
replace the curvature $F$ with $F+B$ throughout. 
Then the BRST-invariance condition is modified by the B-field as follows:
$$
F^{0,2}=-B^{0,2}\ot id_E.
$$
To get a twisted coherent sheaf out of such an object, let us choose a good cover of $X$, so that on any open set
$U_i$ of the cover one has $B^{0,2}=\bpartial \beta_i$ for some $\beta_i\in \Omega^{0,1}(U_i)$. On each $U_i$ let us shift
the covariant $\bpartial$ operator acting on sections of $E\vert_{U_i}$ by $-\ii\beta_i$. The shifted connections 
$\bpartial_i$ satisfy $\bpartial_i^2=0$, so we can define on each $U_i$ a sheaf $S_i$ of holomorphic sections with respect to
$\bpartial_i$. Every $S_i$ is a locally free coherent sheaf on $U_i$. Next, let $\alpha_{ij}$ be a Cech 1-cocycle with
values in $C^\infty(X)$ such that $\beta_i-\beta_j=\bpartial\alpha_{ij}$. We define a sheaf isomorphism 
$$
\phi_{ji}:S_i\vert_{U_{ij}}\ra S_j\vert_{U_{ij}}
$$
by letting
$$
\phi_{ji}:s\mapsto s\exp(\ii\alpha_{ij}),\quad \forall s\in S_i(U_{ij}).
$$
Then the conditions (i)-(iii) are satisfied, provided one takes $\lambda$ to be the Cech 2-cocycle representing the
class of $B^{0,2}$ in $H^2(\O)$.

The remaining piece of the Hochschild cohomology in degree 2 is $H^0(\Lambda^2 T)$. 
Its elements can be represented by holomorphic 
bi-vectors. Given a holomorphic bi-vector $\theta$ and an affine chart $U\subset X$ the infinitesimal deformation of 
the algebra $\O(U)$ in the direction $\theta$ replaces the usual product by
$$
f\star g=fg+\frac{\hp\ii}{2}\theta(df,dg)+O(\hp^2).
$$
Here we introduced the Planck constant $\hp$ to emphasize that the above formula is correct only to first order
in $\theta$. In fact, if one tries to keep terms
of order $\hp^2$, one finds an obstruction: the star-product can be made associative only if $\theta$ is
a Poisson bi-vector. M.~Kontsevich showed that there are no other obstructions to deforming the algebra $O(U)$
up to an arbitrarily high order in $\hp$~\cite{Konts1,Konts2}. Actually, the above requirements do not determine
higher-order terms in the star-product uniquely. Using this freedom, one can ensure that the deformed algebras $\O(U_i)$
fit into a sheaf of noncommutative algebras on $X$. The derived category of the corresponding noncommutative
complex manifold is the deformation of $D^b(X)$ in the direction $\theta$. 
The main goal of this paper is to understand this deformation from a physical viewpoint.

It turns out that this problem is closely related to a certain puzzle which arises when considering T-duality
transformations for $N=2$ SCFTs on flat tori. 
The puzzle is the following. It is well-known that given a torus $T$ with a flat metric $G$ and a flat B-field $B$,
there is a metric $\hG$ and the B-field $\hB$ on the dual torus $\hT$ such that the corresponding $N=1$ SCFTs are isomorphic.
The corresponding triples $(T,G,B)$ and $(\hT,\hG,\hB)$ are called T-dual. The situation becomes rather different 
if we consider $N=2$ SCFTs.
Suppose we have specified a complex structure $I$ on $T$ such that $\omega=GI$ is a symplectic form. This allows
one to enlarge $N=1$ super-Virasoro to $N=2$ super-Virasoro symmetry. Now we ask how $I$ transforms under T-duality.
Somewhat surprisingly, if $B^{0,2}\neq 0$, then {\it there is no choice of complex structure on $\hT$ which makes the two $N=2$
SCFTs isomorphic!} This follows from the results of Ref.~\cite{KO} which we recall below.
The disappearance of T-duality is certainly disturbing, and one would like to restore it in some way. 

The relation of this puzzle to our original problem is the following. From the mathematical viewpoint, 
T-duality acts on B-branes via the Fourier-Mukai transform. As we will see in the next section, if $B^{0,2}\neq 0$,
then the Fourier-Mukai transform relates B-branes on $T$ with B-branes on a noncommutative deformation of $\hT$.
Hence if we can figure out how T-duality works for $B^{0,2}\neq 0$, we will also find out how noncommutativity
is realized on the sigma-model level.

Following this route, we discover that in order to get a noncommutative target-space, one has to to take 
different complex structure for right-movers and left-movers on the world-sheet. This is not an intuitively obvious result,
and to make the argument more convincing we will perform several checks, mostly on the classical
level. For example, we will see that when the left- and right-moving complex structures are different, 
B-branes satisfy a kind of ``uncertainty principle'', as one would expect when the target space is noncommutative.
In the case of tori we will also see how T-duality transformations act on 
the space of various deformations of the category of B-branes.

Our results have several interesting consequences. First of all, we will see that for hyperk\"ahler $X$
the category of B-branes can be continuously deformed into the category of A-branes by turning on noncommutativity
and the B-field. This provides some evidence that A-branes can also be described in terms of modules over a
noncommutative algebra.

Second, we will see that for unequal left-moving and right-moving complex structures it is more natural to describe
the geometry of topological D-branes in terms of generalized complex structures introduced by N.~Hitchin, than in terms
of complex or symplectic geometry. For example, if the bundle on a D-brane is flat, then the D-brane is a generalized
complex submanifold of the Calabi-Yau. We will argue that generalized complex manifolds are in some sense a semi-classical 
approximation to noncommutative complex manifolds with B-fields.

Third, our results suggest a ``practical'' way of constructing the derived category of coherent sheaves on a noncommutative
Calabi-Yau manifold $X$.
It is the category of topological D-branes for the topologically-twisted quantum sigma-model with unequal left-
and right-moving complex structures and a particular B-field. Of course, the usefulness of this definition depends on 
whether one can quantize the sigma-model. In the commutative case, the path-integral of the B-model localizes on constant maps 
to $X$, and there are no quantum corrections. We will see that for unequal complex structures the situation is much like
in the A-model, i.e. the path-integral {\it does not} localize on constant maps, in general, and quantum effects 
can be non-trivial.

In this paper we also derive the condition on the charge vectors of topological D-branes. In general D-brane charge takes
values in $H^*(X,\QQ)$.\footnote{We disregard torsion phenomena in this paper. If one wants to take torsion into
account, then the correct statement is that the charge takes values in $K(X)$~\cite{MM,W}.} It is equal to the Chern character
of the brane times the square root of $\hA(TX)$.
It is well-known that for B-branes with $B^{0,2}=0$ the 
Chern character (and therefore the charge vector) lies in the sub-space
$$
\op_p H^{p,p}(X)\cap H^*(X,\QQ).
$$
There is a simple generalization of this condition to the case when the left- and right-moving complex structures are
unequal. This also includes the case of A-branes, since the A-model can be thought of as a B-model with
the left and right-moving complex structures being opposite.  
We show that for A-branes in the absence of the B-field the condition on the charge vector $v\in H^*(X,\QQ)$ reads:
$$
\left(\frac{\omega}{2\pi} \wedge\, - \,\,\i_{(\omega/2\pi)^{-1}}\right) v=0.
$$
Here $\i_{(\omega/2\pi)^{-1}}$ is
an operator of contraction with the bi-vector $\left(\frac{\om}{2\pi}\right)^{-1}$. We are using the convention $2\pi\alpha'=1$.

The organization of the paper is the following. In section~\ref{FM} we explain why the Fourier-Mukai-dual of a
commutative torus with $B^{0,2}\neq 0$ is a noncommutative complex torus. In section~\ref{Tduality} we analyze
T-duality in the case $B^{0,2}\neq 0$ and show that the T-dual description involves unequal complex structures for
left- and right-movers. In section~\ref{proposal} we formulate our proposal about the sigma-model realization
of the noncommutative deformations. In sections~\ref{test1} and \ref{test2} we perform some tests of the proposal.
In section~\ref{genc} we explain how the language of generalized complex structures can be used to describe 
the geometry of topological D-branes (on the classical level).  In section~\ref{genK} we briefly discuss topologically 
twisted
sigma-models in the case when the left- and right-moving complex structures are unequal and show that
the path integral localizes on generalized holomorphic maps. In section~\ref{Chern} we derive the condition on the
charge vectors of topological D-branes.
In section~\ref{concl} we discuss our results.

A few words about our notations and conventions. To any bi-linear form $Q$ on a vector space $V$ one can assign
a map from $V$ to $V^*$. We will denote this map by the same letter $Q$; thus given $v\in V$ the expression $Qv$
will be an element of $V^*$. On the other hand, an expression $Q(v_1,v_2)$ will denote the value of the bi-linear
form $Q$ on vectors $v_1,v_2$. The natural pairing between $V$ and $V^*$ will be denoted simply as $(v,\xi)$,
or $(\xi,v)$, where $v\in V$ and $\xi\in V^*$. Finally, throughout this paper we let $2\pi \alpha'=1$, unless
stated otherwise. 

{\em Note.} After this paper was posted on the arXiv, I have learned that toroidal $N=2$ sigma-models with unequal
left and right-moving complex structures have been studied in Ref.~\cite{Encke}. In particular, there is
a substantial overlap between Section 3 of this paper and the results of Ref.~\cite{Encke}.

\section{Fourier-Mukai transform and noncommutative complex tori}\label{FM}

Let $V$ be an $n$-dimensional complex vector space, $\Gamma$ be a maximal rank lattice in $V$, and
$T=V/{\sqrt{2\pi}}\Gamma$ be the corresponding complex torus. The dual torus $\hT$ is defined to be
$\hT=V^*/{\sqrt{2\pi}}\Gamma^*$. The factors $\sqrt{2\pi}$ have been inserted for future convenience.
If $I\in \End(V_\RR)$ is a complex structure tensor on $T$, then the complex structure tensor on $\hT$
is $-I^t$, where the superscript $t$ means ``transposed.''

T-duality between $T$ and $\hT$ implies that the categories of B-branes on $T$ and $\hT$ are equivalent.
From the mathematical point of view the equivalence between $D^b(T)$ and $D^b(\hT)$ is explained using
the Fourier-Mukai transform.
In this section we describe what happens to the Fourier-Mukai transform when we turn on a B-field on $T$ which has a
$(0,2)$ piece. This will help us to answer the questions raised in the introduction.

Let us recall how the Fourier-Mukai transform works for $B^{0,2}=0$. The key role is played
by the Poincare line bundle $\cP$ on the product $T\times \hT$. This line bundle has a canonical connection, which, when
lifted to the universal cover of $T\times\hT$, has the form
$$
\nabla=d-\ii A=d-\ii\sum_{i=1}^{2n} \hx_i dx^i.
$$
Here we chose an integral basis for $\Gamma$ and denoted by $x^i$ and $\hx_i$ the corresponding affine coordinates
on the universal cover of $T$  and $\hT$.

We give $\hT$ the dual complex structure. This means that if $I$ is the complex structure tensor for $T$,
then the complex structure tensor for $\hT$ is $-I^t$. 
It is easy to see that the curvature 2-form
$$
F=dA= \sum_{i=1}^{2n} d\hx_i\wedge dx^i
$$
is of type $(1,1)$ in this complex structure, so $\cP$ is a holomorphic line bundle on $T\times\hT$. 

Let $\pi$ and $\hat\pi$ denote the projections from $T\times\hT$ onto $T$ and $\hT$.
The Fourier-Mukai transform of an object
$E\in D^b(T)$ is defined as
$$
{\bf R}{\hat\pi}_*(\cP\ot\pi^* E).
$$ 
That is, one pulls $E$ from $T$ to $T\times \hT$, tensors with $P$, and then pushes
forward to $\hT$. The pull-back, push-forward, and tensor product are understood in the derived sense. It is
crucial for this construction that the Poincare line bundle can be regarded as an object of $D^b(T\times\hT).$

Now suppose that $B^{0,2}\neq 0$ on $T$. Instead of coherent sheaves, we are now dealing with twisted
coherent sheaves. The bounded derived category of $B$-twisted coherent sheaves on $T$ will be denoted $D^b(T,B)$.

If one tries to define the Fourier-Mukai transform for $D^b(T,B)$, one notes that the dual torus $\hT$ does not
carry any natural B-field. Therefore in order for the push-forward from $T\times\hT$ to $\hT$ to be defined,
the object on $T\times\hT$ must live in the ordinary derived category $D^b(T\times \hT)$, not the twisted one.
On the other hand, pulling back an object of $D^b(T,B)$ from $T$ to $T\times\hT$ gives an object of the twisted
derived category $D^b(T\times\hT,\pi^*B)$. Thus we must do something to the Poincare line
bundle in order to make it into an object of $D^b(T\times \hT,-\pi^*B)$. This means that its curvature 2-form must
satisfy
$$
F^{0,2}=\pi^*(B^{0,2}).
$$
This is easy to arrange: one just needs
to make the coordinates $\hx_i$ noncommutative, to wit
\begin{equation}\label{Weyl}
[\hx_i,\hx_j]=i\htheta_{ij}.
\end{equation}
Then the curvature becomes
$$
F=d\hx_i\wedge dx^i-\frac{i}{2}[\hx_i,\hx_j] dx^i\wedge dx^j
=d\hx_i\wedge dx^i+\frac{1}{2}\htheta_{ij} dx^i\wedge dx^j.
$$
If we let
$$
\htheta_{ij}=B_{ij},
$$
we get the desired result. 

Actually, the $(1,1)$ component of $\htheta$ is not constrained by the requirement $F^{0,2}=B^{0,2}$
and can be changed at will. This is to be expected: the $(1,1)$ part of $\htheta$ controls the
commutation relations between holomorphic and anti-holomorphic coordinates on $\hT$, and these are
irrelevant as far as coherent sheaves are concerned.

We conclude that the Fourier-Mukai transform takes objects of $D^b(T,B)$ to objects of the derived category
of coherent sheaves on a noncommutative torus $\hT_\htheta$, where the noncommutativity is of the Moyal type.
We will denote the latter category $D^b_{NC}(\hT_\htheta)$.
In a similar way one can define the Fourier-Mukai transform which goes in the opposite direction. It is
plausible that these transforms are inverse to each other, and therefore $D^b(T,B)$ is equivalent to
$D^b_{NC}(\hT_\htheta).$\footnote{This was also conjectured by D.~Orlov.}

\section{T-duality and $N=2$ supersymmetry}\label{Tduality}

In this section we study the action of T-duality on $N=2$ superconformal structure of the world-sheet theory.
This question has been studied previously in Ref.~\cite{KO}, and we begin by recalling the relevant
results.
Let $T=V/{\sqrt{2\pi}}\Gamma$ be a torus of real dimension $2n$ equipped with a flat Riemannian metric $G\in {\rm Sym}^2(V^*)$,
a flat B-field $B\in \Lambda^2 V^*$, and a constant K\"ahler form $\omega\in \Lambda^2 V^*$. We also define
a complex structure $I\in \End(V)$ by letting $I=G^{-1} \omega$. Given these data, one can define an $N=2$ SCFT 
by quantizing the supersymmetric sigma-model. The right-moving super-currents and the R-current are
\begin{align}
G_+ & =\frac{1}{2}\left(\ii G_{ij}+\omega_{ij}\right)\psi_+^i\partial_+ \phi^j,\no\\
\bG_+ & =\frac{1}{2}\left(\ii G_{ij}-\omega_{ij}\right)\psi_+^i\partial_+ \phi^j,\no\\
J_+ & =-\frac{\ii}{2}\om_{ij}\psi_+^i\psi_+^j.\no
\end{align}
while the left-moving ones are
\begin{align}
G_-&=\frac{1}{2}\left(\ii G_{ij}+\omega_{ij}\right)\psi_-^i\partial_- \phi^j,\no\\
\bG_-&=\frac{1}{2}\left(\ii G_{ij}-\omega_{ij}\right)\psi_-^i\partial_- \phi^j,\no\\
J_-&=-\frac{\ii}{2}\omega_{ij}\psi_-^i\psi_-^j.\no
\end{align}
Our conventions are slightly different from those in Ref.~\cite{KO}.
Note that unlike in Ref.~\cite{KO} we distinguish right- and left-movers with subscripts $\pm$. 

Next let us recall the criteria which ensure that two such SCFTs are isomorphic~\cite{KO}. If we care only about 
$N=1$ superconformal structure, then the answer
is the following: suppose we have two triples $(T=V/{\sqrt{2\pi}}\Gamma,G,B)$ and $(T'=V'/{\sqrt{2\pi}}\Gamma',G',B').$ 
Given these data one can define Euclidean metrics on $V\op V^*$ and $V'\op V'^*$:
\begin{align}
\cG&=\begin{pmatrix}  
G-BG^{-1}B & BG^{-1} \\ -G^{-1}B & G^{-1} 
\end{pmatrix},\no\\
\cG'&= \begin{pmatrix}
G'-B'G'^{-1}B' & B'G'^{-1} \\ -G'^{-1}B' & G'^{-1} 
\end{pmatrix}.\no
\end{align}
These vector spaces also carry canonically defined pseudo-Euclidean metrics of signature $(2n,2n)$:
$$
q=\begin{pmatrix} 0 & 1 \\ 1 & 0\end{pmatrix},\quad
q'=\begin{pmatrix} 0 & 1 \\ 1 & 0\end{pmatrix}.
$$
The two $N=1$ SCFTs are isomorphic if and only if there exists an isomorphism of lattices $\Gamma\op\Gamma^*$
and $\Gamma'\op\Gamma'^*$ which takes $q$ to $q'$ and $\cG$ to $\cG'$. The ``if'' part of this statement
can be formulated in a more familiar form as follows. Let us choose some identification of $\Gamma$ and $\Gamma'$ and $V$ and
$V'$. Consider any linear transformation $f$ of $\Gamma\op\Gamma^*$ which preserves $q$, i.e. belongs
to $O(2n,2n,\ZZ)$. Then given $G,B$ one can construct $G',B'$ giving the same $N=1$ SCFT by letting
$$
\cG= f^t\cG' f.
$$
The simplest example of $f$ is 
$$
f=\begin{pmatrix} 0 & 1 \\ 1 & 0\end{pmatrix}.
$$
This may be called ``T-duality in all directions.'' The formulas for $G'$ and $B'$ take the following simple
form:
\begin{align}\label{TallGB}
G'&=(G-BG^{-1}B)^{-1}=(G+B)^{-1}G(G-B)^{-1},\\
B'&=-G^{-1}B(G-BG^{-1}B)^{-1}=-(G+B)^{-1}B(G-B)^{-1},
\end{align}
or equivalently
$$
G'+B'=(G+B)^{-1}.
$$

We are more interested in isomorphisms of SCFTs which preserve $N=2$ superconformal structure. 
The criterion of isomorphism for toroidal $N=2$ SCFTs looks as follows~\cite{KO}. Instead of $\cG$, we define a pair of complex
structures $\cI$ and $\cJ$ on $V\op V^*$:
\begin{align}
\cI&=\begin{pmatrix} I & 0 \\ BI+I^tB & -I^t\end{pmatrix},\no\\
\cJ&=\begin{pmatrix} \omega^{-1}B & -\omega^{-1} \\ \omega+B\omega^{-1}B & -B\omega^{-1}\end{pmatrix}.\no
\end{align}
One can easily check that $\cG=q\cI\cJ$, so we can recover $G$ and $B$ from $\cI$ and $\cJ$. Similarly,
we define complex structures $\cI'$ and $\cJ'$ on $V'\op V'^*$. The two $N=2$ SCFTs are isomorphic if and only if there
is an isomorphism of lattices $\Gamma\op\Gamma^*$ and $\Gamma'\op\Gamma'^*$ which takes $q$ to $q'$, $\cI$ to
$\cI'$ and $\cJ$ to $\cJ'$. 

Let us consider again ``T-duality in all directions.'' We see that in order for the dual torus to produce
the same $N=2$ SCFT as the original one, the complex structure $\cI'$ must be
$$
\cI'=\begin{pmatrix} -I^t & BI+I^tB \\ 0 & I\end{pmatrix}.
$$
But if $BI+I^t B\neq 0$, this is impossible, because $\cI'$ is always block-lower-triangular, by definition!
Thus it is impossible to find a complex structure $I'$ on the dual torus which gives an isomorphic $N=2$ SCFT.

Note that $BI+I^t B=2\ii(B^{2,0}-B^{0,2})$. Thus problems with T-duality arise precisely when $B^{0,2}\neq 0$,
in which case the category of B-branes is ``twisted'' by $B^{0,2}$.

To understand what is going on, let us compute the image of the generators of $N=2$ super-Virasoro
under T-duality in all directions. A short computation  gives
\begin{align}
G'_+ & =\frac{1}{2}\left(\ii G'_{ij}+\omega'_{+ij}\right){\psi'}_+^i\partial_+ \phi'^j,  \no\\
\bG'_+&=\frac{1}{2}\left(\ii G'_{ij}-\omega'_{+ij}\right){\psi'}_+^i\partial_+ \phi'^j, \no\\
J'_+&= -\frac{\ii}{2}\om'_{+ij}{\psi'}_+^i{\psi'}_+^j,    \no\\
G'_-&= \frac{1}{2}\left(\ii G'_{ij}+\omega'_{-ij}\right){\psi'}_-^i\partial_- \phi'^j,    \no\\
\bG'_-&= \frac{1}{2}\left(\ii G'_{ij}-\omega'_{-ij}\right){\psi'}_-^i\partial_- \phi'^j   \no \\
J'_-&= -\frac{\ii}{2}\om'_{-ij}{\psi'}_-^i{\psi'}_-^j .  \no
\end{align}
Here $G'$ is is given by Eq.~(\ref{TallGB}), while $\omega'_{\pm}$ are given by
$$
\omega'_{\pm}=-(1\pm G^{-1}B)^{-1}\omega^{-1}(1\mp BG^{-1})^{-1}.
$$
We see that the formulas for SUSY generators are the same as before, except that we have
different symplectic forms for right-movers and left-movers.
This explains why we had problems with T-duality in all directions: we were unnecessarily restrictive
in defining generators of $N=2$ super-Virasoro when we assumed that the right-moving and left-moving
symplectic and complex structures are the same. 

In fact, it is well-known that the most general
$N=(2,2)$ supersymmetric sigma-model requires the following data on the target space:
\begin{itemize}
\item[(i)] a Riemannian metric $G$,
\item[(ii)] a 2-form B (not necessarily closed),\footnote{We are being slightly imprecise here. In general, $B$
is not a 2-form, but a connection on a $U(1)$ gerbe. In other words, $B$ is a 2-form only locally,
so $H=dB$ may represent a non-trivial class in de Rham cohomology.}
\item[(iii)] a pair of complex structures $I_+$ and $I_-$ such that $G$ is of type $(1,1)$ with respect to both,
and $I_\pm$ are parallel with respect to the connections
$$
\nabla_\pm=\nabla_0\pm T,
$$
where $\nabla_0$ is the Levi-Civita connection for $G$, and the torsion tensor $T$ is given by 
$$T^i_{jk}=g^{il}(dB)_{ljk}.$$
\end{itemize}
When $dB=0$, one usually takes $I_+=I_-$, 
but this is not necessary to do, and in fact we have seen that such a choice is not preserved by T-duality.

Allowing unequal complex structures for left and right movers, we can repeat the arguments of Ref.~\cite{KO}
and get the criterion for two $N=2$ SCFTs to be isomorphic. To formulate it, we define two commuting complex structures
$\cI$ and $\cJ$ on $V\op V^*$ as follows:
\begin{align}\label{cI}
\cI&=\begin{pmatrix}\tI +(\delta P)B & -\delta P \\
\delta\om+B(\delta P)B+B\tI +\tI^tB & -\tI^t-B\delta P
\end{pmatrix},\\ \label{cJ}
\cJ&=\begin{pmatrix} \delta I +\tP B & -\tP \\
\tom+B\tP B+B\delta I +(\delta I^t)B & -\delta I^t-B\tP.
\end{pmatrix}.
\end{align}
Here we denoted
\begin{align}
\tI&=\frac{1}{2}(I_++I_-),& \delta I&=\frac{1}{2}(I_+-I_-),\no\\
\tom&=\frac{1}{2}(\om_++\om_-), & \delta\om&=\frac{1}{2}(\om_+-\om_-),\no\\
\tP& =\frac{1}{2}(\om_+^{-1}+\om_-^{-1}), & \delta P&=\frac{1}{2}(\om_+^{-1}-\om_-^{-1}),\no\\
\omega_\pm &=G I_\pm. & &\no
\end{align}
Note that $\cG=q\cI\cJ,$ as before.
We define complex structures $\cI'$ and $\cJ'$ on $V'\op V'^*$ in a similar way. The corresponding toroidal
$N=2$ SCFTs are isomorphic if and only if there exists an isomorphism $f$ from $\Gamma\op\Gamma^*$ to $\Gamma'\op\Gamma'^*$
which takes $q$ to $q'$, $\cI$ to $\cI'$, and $\cJ$ to $\cJ'$. 

The above formulas for $\cI$ and $\cJ$ have a simple meaning. The two complex structures $I_+$ and $I_-$ act on $\psi_+$
and $\psi_-$, respectively. In many situations it is more convenient to work with their linear combinations
$$
\psi^i=\frac{1}{2}(\psi_+^i+\psi_-^i),\quad \rho_i=\frac{1}{2}G_{ij}(\psi_+^j-\psi_-^j).
$$
The fields $\psi^i$ can be thought of as components of a single field $\psi$ taking values in the pull-back of $TX$
to the world-sheet. Similarly, $\rho_i$ can be thought of as components of a fermi-field with values in the pull-back of
$TX^*$. Starting with a complex structure
$$
\begin{pmatrix} I_+ & 0\\ 0 & I_-\end{pmatrix}
$$
written in the $\psi_\pm$ basis and writing it in the $(\psi,\rho)$ basis, we get the matrix
$$
\begin{pmatrix} I & -\delta P \\ \delta\om & -I^t \end{pmatrix}.
$$
This is precisely $\cI$ in the special case of vanishing B-field. Similarly, if we start with the complex
structure
$$
\begin{pmatrix} I_+ & 0\\ 0 & -I_-\end{pmatrix}
$$
in the $\psi_\pm$ basis and write it in the $(\psi,\rho)$ basis, we get $\cJ$, in the special case $B=0$.
To get $\cI$ and $\cJ$ with $B\neq 0$, one has to perform an additional basis transformation using
the matrix
$$
\begin{pmatrix} 1 & 0\\ -B & 1 \end{pmatrix}.
$$

Suppose we start with $I_+=I_-$ and perform a T-duality in all directions. Then $\cI'$ must
be upper-triangular:
$$
\cI'=\begin{pmatrix} -I^t & BI+I^tB \\ 0 & I \end{pmatrix}.
$$
This implies that the two complex structures on $\hT$ are given by
\begin{equation}\no
I'_+=-I^t+(BI+I^t B)(G+B)^{-1},\quad I'_-=-I^t-(BI+I^tB)(G-B)^{-1}.
\end{equation}
We see that $I'_+=I'_-$ if and only if $B$ is of type $(1,1)$. 

As explained in the previous section, we expect that for $B^{0,2}\neq 0$ T-duality relates B-branes on $T$
with B-branes on the noncommutative deformation of $\hT$, such that the Poisson bi-vector $\htheta$ on $\hT$
controlling the noncommutativity is 
$$
\htheta=B.
$$
Combining this observation with the results of this section, we conclude that noncommutative deformation of
$\hT$ corresponds to making the left- and right-moving complex structures unequal.

We note that the relation $\htheta=B$ can also be argued on physical grounds. It was shown in Ref.~\cite{SW} that 
if the point-splitting regularization of the path-integral is used, then the connection on a D-brane should be 
regarded as a function
of non-commuting variables with the commutation relations Eq.~(\ref{Weyl}). The bi-vector $\theta$ is given
by\footnote{To compare with Ref.~\cite{SW}, keep in mind that we use the convention $2\pi\alpha'=1$.}:
$$
\theta=-(G+B)^{-1}B(G-B)^{-1}.
$$
But this is precisely the B-field on the dual torus $\hT$ (see Eq.~(\ref{TallGB})), hence $\theta={\hat B}$. 
Exchanging $T$ and $\hT$, we again get $\htheta=B$.

\section{The proposal}\label{proposal}

Now we extrapolate these observations to more general Calabi-Yau manifolds. That is, we propose
that turning on deformations both in $H^2(\O)$ and $H^0(\Lambda^2\Thol)$ is realized on the level of the sigma-model
by allowing generic B-field, as well as unequal left- and right-moving complex structures. 

For this proposal to make sense, any Calabi-Yau which admits noncommutative deformations must also admit
continuous deformations of the complex structure for a fixed metric. To see that this is indeed the case,
recall that $\dim H^0(\Lambda^2 \Thol)=h^{n-2,0}=h^{2,0}$. Thus if a Calabi-Yau admits a noncommutative
deformation, then $h^{2,0}\neq 0$, and by Bochner's theorem~\cite{Bochner} its holonomy group is 
strictly smaller than $SU(n)$. It follows that either $X$ is locally irreducible
and hyperk\"ahler, or it is locally isometric to a product of Calabi-Yaus of smaller dimension. In the latter case,
$X$ is locally a product of several genuine Calabi-Yau manifolds (with exactly one covariantly constant spinor),
some locally irreducible hyperk\"ahler manifold $Y$, and some torus $T$, and it is easy to see that the 2-form and the 
bi-vector ``live'' entirely
on $Y\times T$. Therefore if $X$ admits a noncommutative deformation, then it has at least a 2-parameter 
family of K\"ahler structures for a fixed metric.\footnote{If $X$ is a hyperk\"ahler manifold, or a torus, or 
a product of such manifolds, then any complex structure
can be continuously deformed to minus itself. Now consider a deformation of $N=2$ SCFT where $I_+$ stays fixed, while
$I_-$ starts out equal to $I_+$ and is deformed to $-I_+$. Clearly, the end result is the mirror of the original SCFT.
In particular, this means that in this case the category of B-branes can be continuously deformed into the
category of A-branes by turning on deformations in $H^2(\O)$ and $H^0(\Lambda^2 T).$}

In the next two sections we will perform some tests of this proposal. It will be important for these tests to know
precisely the locus in the moduli space of $N=2$ sigma-models where only the noncommutative deformations
are turned on, while the ``twistedness'' parametrized by $H^2(\O)$ is zero. This is because we do not have an
{\it a priori} understanding of the geometry of B-branes when both kinds of deformations are turned on. (We will return
to this issue in section~\ref{genc}, where we will see that in general the geometry of B-branes is described
using the notion of a ``generalized complex structure.'') In the case of tori, the answer is fairly obvious. 
We know already that the situation where
only ``twistedness'' is turned on is characterized by the fact that the matrix $\cI$ is block-lower-triangular.
Arguments based on Fourier-Mukai transform tell us that after T-duality in all directions only noncommutative
deformations are turned on. Hence purely noncommutative deformations correspond to $\cI$ being block-upper-triangular.
The constraint that $\cI$ is block-upper-triangular translates into a rather complex-looking condition on $\om_\pm$ and $B$.

We can be more precise about the relation between $\cI$ and noncommutativity. According to section \ref{FM},
the coordinates on $\hT$ satisfy:
$$
[\hx_i,\hx_j]=i\htheta_{ij},
$$
where the constant Poisson bi-vector $\htheta$ is given by
$$
\htheta_{ij}=B_{ij}.
$$
On the other hand, the upper right corner of $\hat{\cI}$ is equal to 
$$
-\delta \hP=BI+I^tB. 
$$
Hence we get
$$
\delta \hP=-\htheta I-I^t \htheta=\hI\htheta+\htheta\hI^t.
$$
Here we took into account that the complex structures on $T$ and $\hT$ are related by $\hI=-I^t$.

Note that this relation does not determine $\theta$ uniquely: it leaves undetermined its $(1,1)$ part.
This is hardly surprising, since for the category of B-branes only the noncommutativity of holomorphic
coordinates matters, and the commutation relation between holomorphic and anti-holomorphic coordinates
are of no consequence. If one requires $\theta$ to be of type $(2,0)+(0,2)$, then there is a unique
solution for the above equation:
$$
\htheta=-\frac{1}{2}\hI\delta \hP.
$$

We propose that in general purely noncommutative deformations are characterized by the fact that the matrix $\cI$
is block-upper-triangular. That is, for any Calabi-Yau manifold $X$ equipped with a metric $G$, a closed B-field $B$, and  
a pair of complex structures $I_\pm$ compatible with $G$ and parallel with respect to the Levi-Civita connection, 
one can define $\cI$ by the same formula as for
tori, and use it to decide whether the category of B-branes is a purely noncommutative deformation of 
the category of B-branes on an ordinary Calabi-Yau. Furthermore, we propose that the complex structure of the
corresponding commutative Calabi-Yau can be read off from the diagonal blocks of $\cI$, while the Poisson bi-vector
parametrizing the deformation can be read off the upper-right corner of $\cI$, i.e.
$$
I=\tI+\delta P B,\quad \theta=-\frac{1}{2}I\delta P.
$$

This conjecture can be argued in several ways. First of all, we will see in the next section that it leads
to a correct semi-classical description of B-branes on a noncommutative complex manifold. Second, although
the definition of $\cI$ looks unmotivated, in fact it has rather remarkable properties on an arbitrary
complex manifold.\footnote{Everything we are going to say about $\cI$ also applies to $\cJ$.}
It is easy to check that $\cI$ squares to $-1$, and that it satisfies
$$
\cI^t q \cI=q.
$$
That is, if we regard $\cI$ as a complex structure on the vector bundle $E=TX\op T^*X$, then the pseudo-Euclidean
metric $q$ on $E$ has type $(1,1)$. In the case when $\cI$ is block-upper-triangular, these properties imply
that the diagonal block $I=\tI+\delta P B$ is an almost complex structure on $X$, while the upper-right-corner block 
$-\delta P$ is a bi-vector of type $(2,0)+(0,2)$. In fact, since $\delta P$ is a difference of two Poisson structures,
it is a Poisson bi-vector of type $(2,0)+(0,2)$. 

It is less obvious that the almost complex structure $I=\tI+\delta P B$ is 
integrable, and that $\delta P^{2,0}$ is holomorphic with respect to $I$, but we will see in section~\ref{genc} 
that this is true.

\section{Uncertainty principle for topological D-branes}\label{test1}

The first non-trivial test of our proposal involves the geometry of B-branes in the presence of noncommutativity. In this
section we will assume for simplicity that there is no gauge field on the brane, i.e. the brane
is simply a submanifold in $X$. We will perform all computations is the classical approximation, i.e. to leading
order in the world-sheet Planck constant $\alpha'$. In the limit $\alpha'\ra 0$ the noncommutativity is vanishingly
small. Indeed, if we restore $\alpha'$ in all formulas by replacing
$$
G\ra \frac{1}{2\pi\alpha'} G,\qquad \omega_\pm\ra \frac{1}{2\pi \alpha'} \omega_\pm,
$$
we see that $\delta P$ is of order $\alpha'$. Thus the world-sheet Planck constant $2\pi\alpha'$ can be
identified with the Planck constant $\hbar$ which controls the noncommutativity of space-time coordinates.

Before performing the computation, let us state our expectations for the geometry of B-branes.
As usual, we expect that the support of every B-brane is a complex
manifold $Y$, so locally it is given by equations $f_1=f_2=\ldots=f_k=0$ for some holomorphic functions $f_i$.
If $X$ is noncommutative, there are further constraints on the geometry of $Y$. Indeed, if all $f_i$ vanish on $Y$,
then their commutators must also vanish on $Y$. To first order in noncommutativity, this means that
\begin{equation}
\theta(df_i,df_j)=0\ {\rm on}\ Y\ \forall i,j.\no
\end{equation}
Here $\theta$ is the Poisson bi-vector which controls the noncommutativity. In physics such $Y$ are known as 
first-class constraint surfaces, while in mathematics they are called coisotropic submanifolds of the Poisson manifold 
$X$. Thus we expect that for small $\theta$ (or equivalently, in the classical limit $\alpha'\ra 0$)
B-branes are complex coisotropic submanifolds of $X$. The dimension of any coisotropic submanifold is at least half the 
rank of $\theta$. For example, if $\theta\neq 0$, then points on $X$ are not coisotropic submanifolds and should not correspond to
supersymmetric branes. Thus the requirement of coisotropicity is a kind of ``uncertainty principle'': it puts constraints on 
how well a B-brane can be localized in $X$.

To formulate boundary conditions for sigma-model fields, it is convenient to use linear combinations of the
left-moving and right-moving fermi-fields:
$$
\psi^i=\frac{1}{2}(\psi_+^i+\psi_-^i),\quad \rho_i=\frac{1}{2}G_{ij}(\psi_+^j-\psi_-^j).
$$
The field $\psi$ can be regarded as a section of the pull-back of $TX$ to the world-sheet, while $\rho$
is a section of the pull-back of $TX^*$. The usual boundary conditions on the Fermi-fields say that
on the boundary of the world-sheet $\psi$ lies in $TY\subset TX\vert_Y$, while $\rho+B\psi$ lies in the
conormal bundle $NY^*\subset TX^*\vert_Y$. The boundary conditions on the Bose fields are determined
by $N=1$ supersymmetry, but we will not need their explicit form here.

In order for $Y$ to be a B-brane, it is necessary and sufficient to require that the left-moving and right-moving R-currents
be equal on the boundary, i.e.
$$
\omega_+(\psi_+,\psi_+)=\omega_-(\psi_-,\psi_-).
$$
In terms of $\psi$ and $\rho$ this condition reads:
$$
\delta\om(\psi,\psi)-\delta P(\rho,\rho)+2(\rho, \tI\psi)=0.
$$
It is convenient to express $\rho$ in terms of another Fermi-field $\chi=\rho+B\psi$. The advantage of working
with $\chi$ and $\psi$ is that they are independent fields taking values in $NY^*$ and $TY$, respectively.
In terms of $\chi$ and $\psi$ the current-matching condition is
$$
-\delta P(\chi,\chi)+2 (\chi, (\tI+\delta P B)\psi)=0.
$$
Here we used the condition that the matrix $\cI$ is block-upper-triangular.
It follows that $Y$ must satisfy the following requirements:
\begin{itemize}
\item[(i)] The complex structure $I=\tI+\delta P B$ must preserve $TY\subset TX\vert_Y$;
\item[(ii)] The bi-vector $\delta P$ must vanish when restricted to $NY^*$.
\end{itemize}
The condition (i) says that $Y$ is complex submanifold with respect to $I$. The condition (ii) says that $Y$ is
coisotropic with respect to $P$, and therefore with respect to the $(2,0)+(0,2)$ part of $\theta$.  This is the expected result.

If we repeat this computation for a block-lower-triangular $\cI$, we find the following conditions:
\begin{itemize}
\item[(i$'$)] $I=I_+=I_-$ must preserve $TY$, i.e. $Y$ is a complex submanifold of $X$;
\item[(ii$'$)] $B^{0,2}$ must vanish when restricted to $Y$.
\end{itemize}
These conditions express the fact that $Y$ is a support of a twisted coherent sheaf on $X$.

If $\cI$ is neither upper nor lower triangular, then a B-brane is not a complex submanifold of $X$
but a generalized complex submanifold with respect to the generalized complex structure $\cI$. This will be discussed 
in section~\ref{genc}.

\section{Holomorphic line bundles on noncommutative tori as B-branes}\label{test2}

In this section we consider the case when $X$ is a torus $T$, and the matrix $\cI$ is block-upper-triangular. We will 
focus on B-branes which are line bundles on $X$.
Our goal is to check that the conditions on
the curvature of a connection which ensure $N=2$ supersymmetry can be interpreted as
saying that the line bundle is a holomorphic line bundle on a noncommutative complex torus. 

Suppose our brane is a line bundle $E$ on $T$, with a connection 1-form $A=A_i dx^i$ and
a curvature 2-form $F=dA$. We use the physical convention in which $A$ and $F$ are real forms.
The boundary condition on the Fermi-fields says that $\rho=-(F+B)\psi$ on the boundary. 
The current-matching condition then takes the form
$$
\delta\om(\psi,\psi)+(F+B)\delta P(F+B)(\psi,\psi)+((F+B)\tI+\tI^t(F+B))(\psi,\psi)=0.
$$
Assuming that $\cI$ is block-upper-triangular, we find the following condition on $F$:
\begin{equation}\label{ncbundle}
FI+I^tF=-F\delta P F,
\end{equation}
where $I=\tI+\delta P B$.
If $\delta P=0$, this simply says that $F$ is of type $(1,1)$, i.e. $E$ is a holomorphic line bundle on $T$.
We would like to argue that in general this condition says that $E$ is a holomorphic line bundle on a
noncommutative deformation of $T$.

It has been shown in Ref.~\cite{SW} that on an affine space or a torus there is a change of variables which 
replaces the ordinary
connection 1-form $A$ with a connection 1-form $\hA$ depending on non-commuting coordinates $\hx^i$ 
satisfying
$$
[\hx^i,\hx^j]=i\theta^{ij}.
$$
Here the constant bi-vector $\theta$ is completely arbitrary.\footnote{In this section we will distinguish 
functions of noncommutative variables with a hat. Note that in sections \ref{FM} 
and \ref{proposal} we used a hat to denote objects
living on the dual torus. In this section the dual torus will not appear, so this change of notation should not cause
confusion.} The change of variables from $A$ to $\hA$ is
called the Seiberg-Witten map. We would like to show that if one chooses
$\theta$ so that
\begin{equation}\label{Ptheta}
\delta P=I\theta+\theta I^t,
\end{equation}
then the condition Eq.~(\ref{ncbundle}) is equivalent to the condition that the curvature $\hF$ of the
connection $\hA$ is of type $(1,1)$. 

In general the Seiberg-Witten map is quite complicated,
so we will restrict ourselves to the case when the curvature $F$ is constant. Then the noncommutative
curvature $\hF$ is related to $F$ as follows~\cite{SW}:
$$
\hF=(1+F\theta)^{-1} F.
$$
The condition that $\hF$ is of type $(1,1)$ looks as follows:
\begin{equation}\label{ncbundle2}
(1+F\theta)^{-1} FI+I^t (1+F\theta)^{-1} F=0.
\end{equation}

We would like to check that this condition is equivalent to Eq.~(\ref{ncbundle}). We will do it in two interesting 
special cases. The first case is when $\theta$ is small. We rescale $\theta\ra \hbar\theta$ and
work to first order in $\hbar$. Then Eq.~(\ref{ncbundle2}) becomes
$$
FI+I^t F= \hbar( F\theta FI+I^t F\theta F) +O(\hbar^2).
$$
This is equivalent to
$$
FI+I^t F=-\hbar F(\theta I^t+I\theta)F +O(\hbar^2).
$$
Taking into account Eq.~(\ref{Ptheta}), we get precisely Eq.~(\ref{ncbundle}).

The second special case is when $\theta$ is non-degenerate. Then we can multiply Eq.~(\ref{ncbundle2})
by $\theta$ from the right and get an equivalent condition.
Introducing $X=-F\theta$, this condition can be rewritten as
\begin{equation}\label{eq1}
\frac{X}{1-X} I^t+I^t\frac{X}{1-X}=0.
\end{equation}
On the other hand, upon using Eq.~(\ref{Ptheta}) the condition Eq.~(\ref{ncbundle}) can also be rewritten in terms of the
matrix $X$:
$$
XI^t+I^tX=-2XI^tX.
$$
It is not hard to show that this matrix equation is equivalent to Eq.~(\ref{eq1}).
This means that for non-degenerate $\theta$ Eq.~(\ref{ncbundle}) is equivalent to the requirement that $\hF$ be
of type $(1,1)$, as claimed.

Note that our analysis of the second case did not depend on the assumption that $\theta\sim \delta P$ is small. In other 
words, the result is true to all orders in $\alpha'$. This happens because
constant-curvature connections correspond to conformally-invariant boundary
conditions even on the quantum level.

\section{Generalized complex structures and topological D-branes}\label{genc}

\subsection{Generalized complex structures}

We have seen above that when $\cI$ is either block-upper-triangular or block-lower-triangular,
one can understand the conditions on topological D-branes in terms of either noncommutative or ``twisted''
complex geometry. In this section we show
that the general case can be understood in terms of generalized complex geometry
as defined by N.~Hitchin~\cite{HitchinGC}.

To set the stage, let us recall the definition of a generalized complex structure (GC-structure) on a smooth 
manifold $X$. A generalized complex structure is a bundle map $\cI:TX\op TX^*\ra TX\op TX^*$
satisfying three conditions:
\begin{itemize}
\item[(i)] $\cI^2=-id$;
\item[(ii)] $\cI$ preserves the pseudo-Euclidean metric $q$ on $TX\op T^*X$;
\item[(iii)] $\cI$ is integrable.
\end{itemize}
The last condition requires some explanation. Recall that a sub-bundle of $TX$ is called
integrable if it is closed with respect to the Lie bracket. This notion can be used to define
integrability of an almost complex structure on $X$: an almost complex structure $I:TX\ra TX$
is integrable if and only if the eigenbundle of $I$ with eigenvalue $i$ is an integrable sub-bundle
of $TX_\CC$. Integrability of $\cI$ is defined similarly, except that one replaces $TX$ with $TX\op TX^*$,
and the Lie bracket with the so-called Courant bracket. The Courant bracket is a bilinear operation
on $TX\op TX^*$ defined as follows~\cite{Royt}:
$$
(Y_1+\xi_1)\circ(Y_2+\xi_2) = [Y_1,Y_2]+\Lie_{Y_1}\xi_2-i_{Y_2} d\xi_2,\quad Y_1,Y_2\in TX, \xi_1,\xi_2\in TX^*.
$$
The Courant bracket is not skew-symmetric, but satisfies a kind of Jacobi identity:
$$
a\circ (b\circ c)=(a\circ b)\circ c+b\circ (a\circ c).
$$
Note that the Courant bracket as defined above is skew-symmetric when restricted to any isotropic sub-bundle
of the pseudo-Euclidean bundle $TX\op TX^*$.

Some prefer to use the skew-symmetrized version of the Courant bracket; then the Jacobi identity is violated
by terms which vanish on isotropic sub-bundles of $TX\op TX^*$. For the purposes of this paper, it will be
immaterial which version of the Courant bracket is used.

Any complex structure $I$ gives rise to a generalized complex structure $\cI$ in the following way:
\begin{equation}\label{GCcomp}
\cI=\begin{pmatrix} I & 0 \\ 0 &-I^t \end{pmatrix}.
\end{equation}
More general examples of GC-structures are obtained by noting that any closed 2-form $B$ on $X$ gives rise to
a bundle isomorphism
\begin{equation}\label{fB}
f_B: TX\op TX^*\ra TX\op TX^*,\quad \begin{pmatrix} v \\ \xi \end{pmatrix} \mapsto \begin{pmatrix} 1 & 0 \\ B & 1 \end{pmatrix}
\begin{pmatrix} v \\ \xi \end{pmatrix},
\end{equation}
which preserves the metric $q$ and the Courant bracket. Thus given any GC-structure $\cI$ and a closed 2-form $B$
one can get another GC-structure by applying the above isomorphism. In particular, applying this isomorphism to
the GC-structure Eq.~(\ref{GCcomp}), we obtain:
$$
\cI=\begin{pmatrix} I & 0 \\ BI+I^tB &-I^t \end{pmatrix}.
$$
This is called the B-field transform of Eq.~(\ref{GCcomp}), or, more concisely, the B-field transform of $I$.
It is easy to show that any block-lower-triangular GC-structure is the B-field transform of a complex structure. 

A dual class of examples is provided by GC-structures which are block-upper-triangular. One can show that
any such structure must have the form
$$
\cI=\begin{pmatrix} I & P \\ 0 &-I^t \end{pmatrix},
$$
where $I$ is a complex structure on $X$, and $P$ is a Poisson bi-vector of type $(2,0)+(0,2)$ whose $(2,0)$ part
is holomorphic, and $(0,2)$ part is anti-holomorphic.

A third class of examples of GC structures is provided by symplectic manifolds. A symplectic structure on $X$ gives 
rise to a GC-structure in the following way:
$$
\cI=\begin{pmatrix} 0 & -\omega^{-1} \\ \om & 0\end{pmatrix}.
$$
Of course, the B-field transform of this is again a GC-structure.

Finally, one can show that if $G$ is a Riemannian metric, $I_\pm$ are two complex structures compatible with it
(i.e. $G$ is of type $(1,1)$ with respect to both and the 2-forms $\om_\pm=GI_\pm$ are closed), and $B$ is a closed
2-form, then $\cI$ and $\cJ$ defined by formulas Eq.~(\ref{cI}) and Eq.~(\ref{cJ}) are GC-structures on $X$.
This has been shown by M.~Gualtieri~\cite{Gualt}. This result enables us to fill a hole in the discussion of 
section~\ref{proposal}, where we claimed that if $\cI$ is block-upper-triangular, then $I=\tI+\delta P B$
is an integrable complex structure, and $\delta P$ is a Poisson bi-vector of type $(2,0)+(0,2)$ whose $(2,0)$ part
is holomorphic. It is easy to check that $I^2=-1$, and $\delta P$ is obviously a Poisson bi-vector of type $(2,0)+(0,2)$, 
but why is $I$ integrable, and why is $\delta P^{2,0}$ holomorphic?

To show that this is the case, we use Gualtieri's theorem, which tells us
that the eigenbundle of $\cI$ with eigenvalue $\ii$ is closed with respect to the Courant bracket. Since
$\cI$ is block-upper-triangular, this implies that the $(1,0)$ part of $TX$ with respect to $I$ is also closed with
respect to the Courant bracket, which in this case reduces to the Lie bracket. Therefore $I$ is integrable.

To show that $\delta P^{2,0}$ is holomorphic, note that for a block-upper-triangular $\cI$ the most general
eigenvector with eigenvalue $\ii$ has the form
$$
\begin{pmatrix} u+\delta P\xi \\ -2\ii\xi\end{pmatrix},\quad v\in TX^{1,0}, \xi\in \Omega^{0,1}(X).
$$
Let us consider two special sections of the bundle $\Ker(\cI-\ii)$: the first one will have $\xi=0$ and $u$ holomorphic,
and the second one will have $u=0$ and $\xi=\bpartial f$, where $f$ is an arbitrary anti-holomorphic function.
Their Courant bracket is equal to
$$
[u,\delta P\bpartial f]=u^i(\partial_i (\delta P)^{\bi\bj}) \bpartial_\bj f \bpartial_\bi.
$$
This vector field belongs to $\Ker(\cI-\ii)$ only if it is identically zero. Since $u$ and $f$ are arbitrary,
this means that $\delta P^{0,2}$ is anti-holomorphic, and therefore $\delta P^{2,0}$ is holomorphic.

\subsection{Topological D-branes and generalized complex submanifolds}

Since the formulas for $\cI$ and $\cJ$ fit naturally into the framework of generalized complex structures,
one could expect that the geometry of topological D-branes can also be conveniently formulated in this
language. To do this, we need the notion of a generalized complex submanifold of a GC-manifold $X$ as defined and 
studied by Gualtieri~\cite{Gualt}. 

Let $Y$ be a submanifold of $X$. The tangent bundle of $Y$ is a sub-bundle of $TX\vert_Y$. The normal
bundle $NY$ is the quotient $TX\vert_Y/TY$, while its dual $NY^*$, called the conormal bundle, is
a sub-bundle of $TX^*\vert_Y$. A submanifold $Y$ of $X$ is called a generalized complex submanifold if and only if
the sub-bundle $TY\op NY^*\subset (TX\op TX^*)\vert_Y$ is preserved by $\cI$. Note that a generalized complex
submanifold is not a generalized complex manifold in general. 

In the case when $\cI$ is a B-field transform of a complex structure $I$, a GC-submanifold is a complex submanifold
such that $B^{0,2}\vert_Y=0$~\cite{Gualt}. This is precisely the geometry of a B-brane in the case when $\cI$ in 
Eq.~(\ref{cI}) is block-lower-triangular and the curvature of the connection $F$ on the brane is zero. (For a moment
we will leave aside the question of how to incorporate the effect of non-zero $F$.) Similarly, when
$\cI$ comes from a symplectic structure on $X$, then a generalized complex submanifold is simply a 
Lagrangian submanifold of $X$. This is precisely the geometry of an A-brane in the case when $F=0$.
It is natural to conjecture that in general $Y$ is a topological D-brane if and only if it is a GC-submanifold with 
respect to the GC-structure Eq.~(\ref{cI}). Then the cases of ordinary B-branes and ordinary A-branes correspond
to special cases $I_+=I_-$ and $I_+=-I_-$, respectively.

It is not hard to show that this is indeed the case. We simply repeat the derivation of the conditions on $Y$ as presented
in section~\ref{test1}, but without assuming that $\cI$ is block-upper-triangular. We find that the following must
be true for any pair of Fermi-fields $\chi$ and $\psi$ with values in $NY^*$ and $TY$ respectively:
$$
(\delta\om+B\delta P B+(B\tI+\tI^tB))(\psi,\psi)-\delta P(\chi,\chi)+2(\chi,(\tI+\delta PB)\psi)=0.
$$
This is equivalent to the condition that $\cI$ preserves $TY\op NY^*$. 

To include non-zero $F$ we note that $F$ can enter only in the combination $B+F$. To make use of this information,
we recall that the effect of the B-field can be mimicked by the isomorphism Eq.~(\ref{fB}). Thus, if $\cI$ is
a B-field transform of $\cI_0$, then $Y$ is a generalized complex submanifold with respect to $\cI$
if and only if $\cI_0$ preserves the B-field transform of $TY\op NY^*$. It follows that to include the effect of
$F$, we have to replace $E=TY\op NY^*$ in the definition of a GC-submanifold by a B-field transform of $E$, with
the B-field taken to be $F$. 

Let us state this more formally. A generalized complex brane is a triple $(Y,L,\nabla)$, where 
$Y$ is a submanifold in $X$, $L$ is a Hermitian line bundle on $Y$, and $\nabla$ is a unitary connection on $L$,
such that the following condition is satisfied:
\begin{itemize}
\item Let $E$ be a sub-bundle of $(TX\op TX^*)\vert_Y$ defined by the condition that $(v,\xi)$ belongs
to $E$ if and only if $v\in TY$ and the image of $\xi$ under the projection to $TY^*$ is equal to $Fv$,
where $F$ is the curvature of $\nabla$. Then $E$ must be preserved by $\cI$.
\end{itemize}

It is easy to see that when $F=0$, the notion of a generalized complex brane (GCB) coincides with the notion of a generalized
complex submanifold equipped with a line bundle and a unitary flat connection. From what was said above, it should be clear 
that the triple $(Y,L,\nabla)$ is a GCB if and only if it is a topological D-brane. 
In the cases when $\cI$ is block-upper-triangular
or block-lower-triangular, we recover noncommutative and ``twisted'' B-branes discussed in the previous sections.

As mentioned above, A-branes arise in the special case $I_+=-I_-$. Alternatively, one may keep $I_+=I_-$, but
replace $\cI$ with $\cJ$. It follows that $(Y,L,\nabla)$ is an A-brane if and only if it is a GCB with respect to the 
GC-structure $\cJ$ with $I_+=I_-$. The conditions for $(Y,L,\nabla)$ to be an A-brane have been previously analyzed in 
Ref.~\cite{KO3}. Needless to say, the results of Ref.~\cite{KO3} are in agreement with our more general discussion here.

\section{Generalized K\"ahler structures and topological sigma-models}\label{genK}

\subsection{Generalized K\"ahler structures}

In the preceding section we have seen that the language of generalized complex structures is very
convenient for describing the geometry of topological D-branes. In this section we discuss topologically
twisted sigma-models in the case when $I_+\neq I_-$. Our goal is to generalize the results of Ref.~\cite{Witten},
where the case $I_+=\pm I_-$ was studied. We will see that our results
are naturally formulated using the notion of a {\it generalized K\"ahler structure}. The latter also
enters in the formulation of stability conditions for topological D-branes, a subject which we hope
to discuss elsewhere.

A generalized K\"ahler structure on a smooth manifold $X$ is a pair of commuting GC-structures $\cI$ and
$\cJ$ such that $\cG=q\cI\cJ$ is a positive-definite metric on $TX\op TX^*$. Here $q$ is the usual pseudo-Euclidean
metric on $TX\op TX^*$. The motivation for this
definition is the following. A K\"ahler manifold has two natural GC-structures: one coming from the complex structure,
and one coming from the symplectic form. It is easy to check that these two GC-structures commute and satisfy
$$
q\cI\cJ=\begin{pmatrix} G & 0 \\ 0 & G^{-1}\end{pmatrix}.
$$
Thus a K\"ahler manifold has a natural generalized K\"ahler structure.

The relevance of generalized K\"ahler structures for physics is shown by the following theorem of Gualtieri~\cite{Gualt}
which was already mentioned in section~\ref{genc}:

{\bf Theorem.} The set of generalized K\"ahler structures on $X$ is in one-to-one correspondence with the set of
the following data:
\begin{itemize} 
\item[(i)] A Riemannian metric $G$ and a 2-form $B$ on $X$;
\item[(ii)] A pair of complex structures $I_\pm$ such that $G$ is of type $(1,1)$ with respect to both,
and $I_\pm$ are parallel with respect to the connections $\nabla_\pm=\nabla_0\pm T,$
where $T^i_{jk}=g^{il}(dB)_{ljk}.$
\end{itemize}

It is well-known that the data described by (i)-(ii) classify $N=(2,2)$ supersymmetric sigma-models with target 
$X$~\cite{torsion}.\footnote{We are assuming again that the B-field $B$ is a globally defined 2-form.}
Thus generalized K\"ahler structures provide an equivalent way of describing the geometry of $N=(2,2)$ sigma-models. 

Note that generalized K\"ahler manifolds are not K\"ahler if $H=dB\neq 0$. Indeed, if we let $\omega_\pm=GI_\pm,$,
then $d\omega_\pm=\mp 2 \i_{I_\pm} H$. In this paper we set $H=0$, although the more general case is also interesting.

\subsection{Topological sigma-models}

In this subsection we discuss topologically twisted sigma-models in the case when $I_+\neq I_-$ and $dB=0$.
On a Calabi-Yau $X$ with $h^{0,2}=0$ one has $I_-=\pm I_+$, so there are only two possibilities for twisting, 
which lead to the well-known A- and B-models. Therefore we will be interested in the case when $X$
is a Calabi-Yau with $h^{0,2}\neq 0$. For simplicity we will mostly consider the case of vanishing B-field in this section.

Our goal is to determine the ring of observables in the closed string case. It is well known that
as a vector space it is completely independent of the choice of $I_\pm$ and isomorphic to the de Rham
cohomology of $X$~\cite{LVW}. But the ring structure depends on $I_\pm$ in an interesting way~\cite{Witten}. 
For example, if $I_+=I_-$
(the usual B-model), the ring of observables is given by
$$
\oplus_{p,q} H^p(\Lambda^q \Thol).
$$
On the other hand, for $I_+=-I_-$ (the A-model) the ring of observables is the quantum cohomology ring of $X$,
which is a deformation of the de Rham cohomology ring depending on the symplectic structure on $X$.
The quantum product is defined in terms of counting holomorphic maps from $\PP^1$ to $X$. The difference between the two cases
stems from the fact that for the B-model the path-integral localizes on constant maps to $X$,
while for the A-model is localizes on holomorphic instantons~\cite{Witten}. We will see below that for generic $I_\pm$ the
situation is similar to that for the A-model, except that holomorphic maps are replaced with "generalized 
holomorphic maps."

First, let us determine the space of observables. For simplicity, we will limit ourselves to
the case of constant metric and complex structure, but we expect the results to hold more 
generally. The BRST-charge density looks as follows:
$$
Q=\frac{\ii}{2}G(\partial_+\phi,(1+\ii I_+)\psi_+)+\frac{\ii}{2}G(\partial_-\phi,(1+\ii I_-)\psi_-).
$$
From previous experience, we know that it is more convenient to work with fermi-fields $\psi=\frac{1}{2}(\psi_++\psi_-)$
and $\rho=\frac{1}{2} G(\psi_+-\psi_-)$ which take values in the pull-back of $TX$ and $TX^*$, respectively.
In terms of $\psi$ and $\rho$ BRST-charge density takes the form:
$$
Q=\ii(p,\partial_1 \phi)(1+\ii\cI)\begin{pmatrix} \psi \\ \rho\end{pmatrix},
$$
where $p_i=G_{ij}\partial_0\phi^j$ is the momentum conjugate to $\phi^i$. The complex structure $\cI$
is the same as in Eq.~(\ref{cI}), except that one must set $B=0$. 

From the formula for $Q$ and the anti-commutation relations
$$
\{\psi^i,\psi^j\}=\frac{1}{2}(G^{-1})^{ij},\quad \{\rho_i,\rho_j\}=\frac{1}{2}G_{ij},
$$
it is easy to see that BRST-invariant fermi-fields
are precisely those ones which are annihilated by $1-\ii\cI$. Let us introduce some notation. The bundle
$TX_\CC\op TX^*_\CC$ decomposes into a direct sum of vector bundles $\V\op \bV$, where $\V$ is an eigenbundle
of $\cI$ with eigenvalue $\ii$, and $\bV$ is an eigenbundle with eigenvalue $-\ii$. BRST-invariant fermions ``live''
in $\bV$. Thus we look for BRST-invariant observables which have the following form:
$$
V_f=f_{a_1 \ldots a_k}(\phi) \Psi^{a_1}\ldots \Psi^{a_k},
$$
where $k=0,1,\ldots,\dim \V$, and
$$
\Psi=\frac{1+\ii\cI}{2}\begin{pmatrix} \psi \\ \rho\end{pmatrix}
$$
is a fermi-field taking values in $\bV$. The functions $f_{a_1\ldots a_k}$ are components of a section
of the bundle $\Lambda^k \bV^* \simeq \Lambda^k \V$. The BRST variation of $V_f$ looks as follows:
$$
\delta V_f=2V_{d_\V f},
$$
where $d_\V f$ is a section of $\Lambda^{k+1} \V$ given by
\begin{equation}\label{dE}
(d_\V f)_{a_0\ldots a_{k}}=\partial_{[a_0} f_{a_1,\ldots,a_k]}.
\end{equation}
Here we defined
$$
\partial_{a_0}=\Pi_{a_0}^i\frac{\partial}{\partial\phi^i},
$$
where $\Pi$ is the natural projection from $\bV$ to $TX_\CC$ (the anchor map, in the terminology of the Appendix).
Thus the exterior algebra bundle 
$$
\cA_\V=\op_p\Lambda^p \V
$$
becomes a differential graded vector space with differential $d_\V$, 
and its cohomology is
precisely the BRST cohomology of the topologically twisted sigma-model.

The mathematical meaning of the complex $(\Lambda^*\V,d_\V)$ is explained in the Appendix. Here we just
note that the cohomology of $d_\V$ in degree $2$ parametrizes infinitesimal
deformations of the GC-structure on $X$~\cite{Gualt}. In analogy with the usual B-model~\cite{Witten}, we may regard the space of 
topological observables as the space of ``extended deformations'' of the GC-structure. 

It is easy to see
that in the cases $I_+=I_-$ and $I_+=-I_-$ the algebra $\cA_\V$ reduces to
$$
\oplus_{p,q} \Lambda^q\Thol\otimes \Omega^{0,p}(X)
$$
and 
$$
\op_p \Omega^p(X),
$$
respectively, and the differential $d_\V$ becomes the Dolbeault differential in the first case, and the de Rham
differential in the second case. Therefore in these two special cases we recover the spaces of observables
in the B- and A-models, respectively.

Finally let us discuss briefly the ring structure on the space of observables. Since $(\cA_\V,d_\V)$ is a dg-algebra, its
cohomology has a natural algebra structure. It is easy to see that it coincides with the ``physical'' ring
structure on the BRST cohomology in the classical approximation. But as in Ref.~\cite{Witten}, there could be 
instanton corrections to the
classical ring structure. Thus to compute the BRST ring we need to identify instantons, i.e. field configurations
on which the path-integral localizes. As explained in Ref.~\cite{Witten} these are precisely BRST-invariant
field configurations.
Requiring the BRST-variations of the fermi-fields to vanish, we find an equation on the map $\phi$ from the world-sheet
$\Sigma$ to the target space $X$:
$$
(1+\ii\cI^t)\begin{pmatrix} G\partial_0\phi \\ \partial_1\phi\end{pmatrix}=0.
$$
Obviously this equation does not have nontrivial solutions when the world-sheet has Lorenzian signature, but 
we are actually interested in Euclidean solutions (instantons). Therefore we need to perform the Wick rotation
$$
\partial_0\phi\ra \ii \partial_2\phi.
$$
Then the above equation is equivalent to the following four equations:
\begin{align}\label{genhol1}
\tom\partial_1\phi&=0, & \tom\partial_2\phi &=0,\\ \label{genhol2}
\partial_2\phi & =\delta I \partial_1\phi, & \partial_1\phi &=-\delta I\partial_2\phi.
\end{align}
The first two equations actually follow from the last two.

To understand the meaning of these equations, we note that the differential $d\phi:T\Sigma\to TX$
can be composed with the standard embedding
$$
j: TX\to TX\op TX^*.
$$
On $T\Sigma$ we have a complex structure $I_\Sigma$, while the generalized complex structure $\cJ$
gives rise to a complex structure on the vector bundle $TX\oplus TX^*$. It is easy to check that 
the equations Eqs.~(\ref{genhol1},\ref{genhol2}) are equivalent to the
following condition:
\begin{equation}\label{genholfree}
\cJ (j\circ d\phi)=(j\circ d\phi)I_\Sigma. 
\end{equation}
Here $\cJ$ is ``the other half'' of the generalized K\"ahler structure on $X$ given explicitly by Eq.~(\ref{cJ}).
Clearly, this condition is an analogue of the holomorphic instanton equation, where the role of the complex structure
on $X$ is taken by the generalized complex structure $\cJ$. We will call such maps generalized
holomorphic maps from $\Sigma$ to $X$.

By analogy with Ref.~\cite{Witten} we expect that the ring structure on the cohomology of $\cA_\V$ dictated by physics 
is not the classical one, but involves summation over generalized holomorphic maps from $\Sigma$ to $X$, 
as in the A-model case. In fact, the A-model is a special case of this: if $I_+=-I_-$, then $\cJ$ becomes
a GC-structure coming from a complex structure, and generalized holomorphic maps become ordinary
holomorphic maps. On the other hand, if $I_+=I_-$, then $\cJ$ is a GC-structure coming from a symplectic
structure, and generalized holomorphic maps are simply constant maps. Thus we also recover the well-known
fact that the B-model path integral localizes on constant maps. 

It is interesting to ask what happens if we start with the ordinary B-model ($I_+=I_-$)
and deform it by making $I_+\neq I_-$. The generalized holomorphic map equation requires
$\d\phi$ to be in the kernel of $\tom$. Since initially $\tom=\om$ is non-degenerate, for
sufficiently small deformations $\tom$ will also be non-degenerate, and the only generalized
holomorphic maps are constant ones. For large deformations
$\tom$ may become degenerate, and then non-trivial generalized holomorphic maps may appear.
The extreme case of this is the A-model, where $\tom$ is identically zero.

We would like to stress that although the path-integral of the sigma-model does not localize on
constant maps, in general, the instanton corrections to the ring of observables often vanish. 
For example, consider the case when $X$ is a hyperk\"ahler manifold and $I_+=-I_-$ (the A-model).
It is well known that in this case instanton corrections to the classical ring structure on $H^*(X)$ vanish.
On the other hand, if we consider the same model on a world-sheet with boundaries (i.e. if we consider
topological D-branes on $X$), then instanton effects are non-trivial. This is true even if $X$ is a torus~\cite{PZ}.
By analogy, we expect that generalized holomorphic instantons
have non-trivial effect on the structure of the category of D-branes. In this connection we note
that it would be interesting to construct an analogue of the Floer homology for generalized complex
submanifolds of a generalized K\"ahler manifold.

So far in this section we have considered the case $B=0$. The general case is easily recovered by applying
the map Eq.~(\ref{fB}) everywhere. For example, the space of observables in the topologically twisted theory
will be the cohomology of $d_\V$, where $\V=\Ker (1+\ii\cI)$, and $\cI$ is given by Eq.~(\ref{cI}).
The generalized holomorphic map equation is the again given by Eq.~(\ref{genholfree}), but with
$\cJ$ given by Eq.~(\ref{cJ}) with an arbitrary B-field. In components, the generalized holomorphic
map equation reads:
\begin{gather}
(\tom+B\tP B +B\delta I+(\delta I)^t B)d\phi=0,\no\\
\partial_2 \phi =(\delta I +\tP B)\partial_1\phi,\no\\
\partial_1 \phi =-(\delta I +\tP B)\partial_2\phi.\no
\end{gather}

\section{Constraints on the charge vector of a topological D-brane}\label{Chern}

In this section we derive a condition on the charge vector of a topological D-brane. Consider a D-brane wrapping a 
submanifold $Y$ of the target-space $X$. If $H=dB=0$, as we have assumed, then $Y$ must be a $Spin^c$-manifold~\cite{W,FW}.
Let $d\in H^2(Y,\ZZ)$ denote the cohomology class of the $Spin^c$-structure on $Y$. Further, let $E$ be the vector
bundle of the D-brane. The charge vector of the D-brane $(Y,E)$ is the following class in $H^*(X,\QQ)$~\cite{MM,CheYin}:
$$
v(Y,E)=f_!\left(\ch(E)e^{\frac{d}{2}}{\sqrt\frac{\hA(TY)}{\hA(NY)}}\right).
$$
Here $f_!$ is the Gysin push-forward map in cohomology:
$$
f_!: H^*(Y)\ra H^{*+{\rm codim Y}}(X).
$$
It is a composition of the Poincar\'{e} duality on $Y$, push-forward on homology, and the Poincar\'{e} duality on $X$.
In fact, it is better to think about the D-brane charge as taking value in  $K^*(X)$, but in this paper we will
not worry about torsion phenomena, and will be content with the above definition of the charge.

The expression which appears in the argument of $f_!$ differs by a factor $e^{-B/(2\pi)}$ from the expression $D_{WZ}$ 
which appears in the Wess-Zumino term in the D-brane action:
$$
S_{WZ}\sim \int_Y C\wedge D_{WZ}= \int_Y C\wedge \ch(E) e^{\frac{d}{2}+\frac{B}{2\pi}}{\sqrt\frac{\hA(TY)}{\hA(NY)}}.
$$
Here $C\in\Omega^*(X)$ is the Ramond-Ramond field. The physical origin of this difference was explained in Ref.~\cite{Taylor}
(see also Ref.~\cite{AMM}). Note that the differential form $D_{WZ}$ is invariant under gauge transformations
$$
B\mapsto B+d\Lambda, \quad \nabla_E\mapsto \nabla_E-\ii\Lambda,
$$
while the differential form representing $v(Y,E)$ is not. Of course, the cohomology class $v(Y,E)$ is gauge-invariant.

If the D-brane in question preserves $N=2$ supersymmetry on the world-sheet, then the charge vector satisfies additional
constraints. For example, for ordinary B-branes with $I_+=I_-$ and $B^{0,2}=0$ it is well-known that the charge vector
satisfies
\begin{equation}\label{Hodge}
v(Y,E)\in \op_p H^{p,p}(X).
\end{equation}
Our goal is to find a generalization of this condition to the case $I_+\neq I_-$ and arbitrary B-field. This includes
A-branes as a special case.

From the point of view of supersymmetric sigma-models, the charge vector is related to the zero-mode part of the boundary 
state of the D-brane in the Ramond-Ramond sector. Let us recall what this means. In the zero-mode approximation 
one retains only the constant modes of the 
bosonic field $\phi$ and fermionic fields $\psi_\pm$. Instead of $\psi_\pm$ we will be using their linear combinations
$\psi$ and $\rho$. Since we have switched from the open-string channel to the closed-string channel, the role
of world-sheet time is now played by $\sigma^1$, while $\sigma^0$ becomes space-like. More precisely, we
define the zero-modes of $\psi$ and $\rho$ to be:
$$
\psi_0=\sqrt 2\int_0^{2\pi} \psi(\sigma^0) d\sigma^0,\quad \rho_0=\frac{1}{\pi\sqrt 2}\int_0^{2\pi} \rho(\sigma^0) d\sigma^0.
$$
Numerical factors here are chosen for future convenience. 
Canonical commutation relations
for $\psi_0$ and $\rho_0$ read
$$
\{\psi_0^i,\psi_0^j\}=\{\rho_{0i},\rho_{0j}\}=0,\quad \{\rho_{0i},\psi_0^j\}=-\delta_i^j.
$$
Thus we can regard $\rho_0$ as canonically conjugate to $\psi_0$. The boundary state in the zero-mode approximation
will depend on the zero-modes of $\phi$ and $\psi$. 
If we interpret $\psi_0^i$ as $dx^i$, then the boundary state
becomes a differential form on $X$. This form is the image of $D_{WZ}$ under the Gysin map. To get the form representing
the charge vector $v(Y,E)$, one has to multiply it by $e^{-B/(2\pi)}$.

The constraint of $N=2$ supersymmetry boils down to the R-current matching condition on the boundary
$$
J_+-J_-=0\ {\rm on}\ \partial\Sigma.
$$
Since we have switched to the closed-string channel, this is now interpreted as a requirement that $J_+-J_-$
annihilate the boundary state. In the zero-mode approximation $J_+-J_-$, which is a function of $\phi,\psi,$ and $\rho$,
becomes a degree-0 differential operator on forms. Therefore the constraints of $N=2$ supersymmetry can always
be expressed as the requirement that a certain linear operator of degree 0 annihilates the boundary state. 
All that remains is to compute this operator and then express the constraint in terms of the charge vector 
$v(Y,E)\in H^*(X,\QQ)$.

In terms of $\rho_0$ and $\psi_0$ the zero-mode part of the operator $J_+-J_-$ takes the form
$$
J_{+0}-J_{-0}=\frac{1}{4\pi}\delta\omega(\psi_0,\psi_0)-\pi\delta P(\rho_0,\rho_0)-2(\psi_0,\tI^t\rho_0).
$$
Note that there is no normal-ordering ambiguity in this operator because the endomorphism $\tI$ is traceless.
Canonical commutation relations imply that $\rho_0$ acts as a differential operator
$$
\rho_{0i}=-\frac{\partial}{\partial\psi_0^i}.
$$
Then we get the following 
condition on the boundary state:
$$
\left(\frac{1}{4\pi}\delta\om(\psi_0,\psi_0)-\pi\delta P(\frac{\partial}{\partial \psi_0},\frac{\partial}{\partial \psi_0})
+(\psi_0,\tI^t\frac{\partial}{\partial\psi_0})\right) f_!(D_{WZ})=0.
$$
Identifying $\psi_0^i$ with $dx^i$, we can rewrite it as
$$
\left((2\pi)^{-1}\delta\om\wedge - 2\pi\,\, \i_{\delta P}\, +\, \i_\tI\right) f_!(D_{WZ})=0.
$$
Here $\i_{\delta P}$ and $\i_\tI$ denote the operators of interior multiplication by the bi-vector 
$\delta P$ and the $(1,1)$ tensor $\tI$, respectively. Thus the three terms in brackets in the above equation
have form-degrees $2$, $-2$ and $0$, respectively. Finally, setting 
$$
f_!(D_{WZ})=e^{\frac{B}{2\pi}}v(Y,E),
$$
we get
\begin{multline}
\left((2\pi)^{-1}\left(\delta\om+B\delta P B +BI+I^t B\right)\wedge - 2\pi\,\, \i_{\delta P}\, +\, \i_{\tI+\delta P B}\right.\\
\left. -\left(\i_{\delta P} B\right)\right) v(Y,E)=0.
\end{multline}
This is the final form of the condition on the charge vector. 

Let us look at some special cases. First consider ordinary B-branes, i.e. $\delta\om=0$, $\delta P=0$, $B^{0,2}=0$, 
and $\tI=I$ is a complex structure. Then the condition reads
$$
\i_I v(Y,E)=0.
$$
This is equivalent to the requirement Eq.~(\ref{Hodge}).

Second, consider ordinary A-branes with $B=0$. In this case $I_+=-I_-$, $\om_+=-\om_-=\om$, $\tI=0$, and $\delta P=\om^{-1}$.
Then the condition becomes
$$
\left(\frac{\om}{2\pi}\wedge\, -\,\,\i_{(\om/2\pi)^{-1}}\right) v(Y,E)=0.
$$
It is easy to see that the charge vector of a Lagrangian submanifold which is spin and carries a flat connection satisfies 
this condition. We expect that
coisotropic A-branes of Ref.~\cite{KO} also satisfy this condition, although we have not checked this in full generality.

\section{Discussion}\label{concl}

It is well-known that the most general $N=2$ structure for a supersymmetric sigma-model involves two different complex
structures $I_\pm$, one for the left-movers and one for the right-movers. 
In this paper we have argued that 
such sigma-models provide a physical realization of noncommutative deformations of Calabi-Yau manifolds.
This means that the category of topological D-branes for such sigma-models is a noncommutative deformation 
of the derived category of coherent sheaves. We also found that for $I_+\neq I_-$ the geometry of topological
D-branes can be described in the language of generalized complex structures.

Since for hyperk\"ahler manifolds and tori one can continuously interpolate between $I_+=I_-$ and $I_+=-I_-$,
our observation lends support to speculations (see e.g. Ref.~\cite{BreSoi}) that A-branes on $X$ are related to 
sheaves on a noncommutative deformation of $X$.

An important question which we completely ignored so far is whether noncommutative deformations of the category of
D-branes make sense for finite values of $\hbar$, or only as a formal series in $\hbar$. Since we have identified 
$\hbar$ with $2\pi\alpha'$, this is related to the question whether topological sigma-models on Calabi-Yau manifolds
are well-defined away from the large-volume limit. For an arbitrary Calabi-Yau $X$, this is a hard question.
However, for some interesting $X$ the answer seems to be yes, at least in the closed-string case.
The simplest example is that of a torus with a flat metric and constant B-field, in which case the conformal
field theory is manifestly well-defined. Therefore we expect that for tori noncommutative deformations make
sense for finite $\hbar$.  A much bigger class of examples is provided by complete intersections
in toric varieties. It is known that sigma-models on such manifolds arise as the
infrared limit of gauged linear sigma-models~\cite{Wittenphases}. The latter theories are free in the ultraviolet
and seem to be well-defined on the non-perturbative level. In particular, if we take $X$ to be a quartic K3
surface in $\PP^3$, the supersymmetric sigma-model is likely to be well-defined, and it is plausible that noncommutative
deformations of $D^b(X)$ make sense for finite $\hbar$.

An important open problem is to give a mathematically precise definition of the category of topological
D-branes in the case when $I_+\neq I_-$. It is not obvious how to do this even when $X$ is a torus. 
The results of this paper suggest that given any generalized
complex structure $\cI$ on $X$ one should be able to define D-branes and morphisms between them. The case
which we understand well is when $\cI$ is block-lower-triangular. In this case the category of topological
D-branes is believed to be equivalent to the derived category of twisted coherent sheaves on $X$. 
If $\cI$ is block-upper-triangular, then, as we have argued in this paper, the relevant category is the derived
category of coherent sheaves on a noncommutative deformation of $X$. These two cases are special in that
$\cI$ can be naturally regarded as a deformation of a complex structure $I$, which can be read off the diagonal
blocks of $\cI$. If $\cI$ is neither block-upper-triangular nor block-lower-triangular, then there is
no canonical choice of $I$, and there is no obvious way to relate topological D-branes to coherent sheaves. 
Therefore one has to try to define the category of topological D-branes starting directly from $\cI$.
A solution of this problem would provide, as a special case, a new definition of the category of A-branes. We note that the
difficulties presented by the case $I_+\neq I_-$ are the same as in the special case $I_+=-I_-$ (A-branes)~\cite{KOlect}.
For example, we understand reasonably well the geometry of D-branes which are line
bundles on a submanifold of $X$, but know very little about D-branes of higher rank. In this
connection, the constraint on the charge vector of a D-brane derived in section~\ref{Chern} could be a useful
guide, since it is valid for D-branes of arbitrary rank. 

Even for topological D-branes of rank one, we did not give a precise recipe for computing the ring of boundary observables.
In mathematical terms, we have not described the structure of a category on the set of topological D-branes.
It seems that to solve this problem one has to generalize the construction of Floer homology from Lagrangian submanifolds
in a symplectic manifold to generalized complex submanifolds in a generalized complex manifold. 

In conclusion we note that it would be interesting to study topologically twisted sigma-models and topological
D-branes for supersymmetric backgrounds with $H=dB\neq 0$.

\section*{Appendix: The BRST complex as a deformation complex of a GC-manifold}

Recall that although
the Courant bracket is not a Lie bracket, it becomes a Lie bracket when restricted to any isotropic 
integrable sub-bundle of $TX\op TX^*$. By the definition of the generalized complex structure, the sub-bundle
$$
\bV=\Ker (1-\ii\cI)
$$
is isotropic and integrable, so its space of sections has a natural Lie bracket. 

In fact, $\bV$ is a (complex)
Lie algebroid. This means that there is a Lie-algebra homomorphism $\Pi:\Gamma(\bV)\ra \Gamma(TX_\CC)$, called the anchor,
which satisfies:
$$
[f \xi,\eta]=f\cdot[\xi,\eta]-\Pi(\eta)(f)\cdot\xi,\quad \forall\xi,\eta\in\Gamma(\bV),\quad \forall f\in C^\infty(X).
$$
In our case the anchor map is given by projecting $\bV\subset TX_\CC\op TX^*_\CC$ onto the first summand.

It is well known that all the usual constructions of differential geometry make sense
when one replaces the tangent bundle with an arbitrary Lie algebroid $\cE$ (see e.g. Ref.~\cite{Liealg}). In particular, 
if $\cE$ is an arbitrary Lie algebroid,
the algebra bundle 
$$
\op_p\Lambda^p \cE^*
$$
has a natural differential analogous to the de Rham differential. It is defined using an obvious generalization
of the Cartan formula to the Lie-algebroid case. Namely, if $\omega$ is a section of $\Lambda^k \cE^*,$ then
$d\omega$ is a section of $\Lambda^{k+1}\cE^*$ whose value on $\xi_0,\ldots,\xi_k\in\Gamma(\cE)$ is 
\begin{multline}
(d\omega)(\xi_0,\ldots,\xi_k)=\sum_{i=0}^k (-1)^i \Pi(\xi_i)(\omega(\xi_0,\ldots,\widehat\xi_i,\ldots,\xi_k))\\
+\sum_{i<j} (-1)^{i+j} \omega([\xi_i,\xi_j],\xi_0,\ldots,\widehat\xi_i,\ldots,\widehat\xi_j,\ldots,\xi_k).
\end{multline}

Let us give one example. If $X$ is a complex manifold, then $\cE=TX^{0,1}$ is a complex Lie algebroid, and
the corresponding differential complex is the Dolbeault complex $(\Omega^{0,*},\partial)$. 

Similarly, if $X$ is a generalized complex manifold, and $\cE=\bV$ is an eigenbundle of $\cI$ with eigenvalue
$-\ii$, this construction gives a differential on the algebra
$$
\cA_\V=\Lambda^* \bV^*\simeq \Lambda^* \V.
$$
In the case when the generalized complex structure is constant, one can show that this differential is precisely $d_\V$ as defined in Eq.~(\ref{dE}).

In the case $I_+=I_-=I$ we have
$$
\V=TX_\CC^{1,0}\oplus (TX_\CC^*)^{0,1},
$$
and $\cA_\V$ is simply the algebra 
$$
\oplus_{p,q} \Lambda^q\Thol\otimes \Omega^{0,p}(X)
$$
with the Dolbeault differential. Its cohomology is the space of states of the usual B-model.

In the case $I_+=-I_-$ we have
$$
\V\simeq TX_\CC,
$$
with the usual Lie bracket, and the algebra $\cA_\V$ is the algebra of differential forms on $X$ with the
de Rham differential. Its cohomology is the space of states of the usual A-model.

\section*{Acknowledgments} 
I would like to thank Marco Gualtieri, Nigel Hitchin, and Dmitri Orlov for discussions. I also would like
to thank the organizers of the workshop ``Geometry and Topology of Strings'' at KITP, UC Santa Barbara, July-August 2004,
for a very stimulating meeting. This research
was supported in part by the DOE grant DE-FG03-92-ER40701.

\end{document}